\documentclass[fleqn,usenatbib]{mnras}

\usepackage{newtxtext,newtxmath}

\usepackage[T1]{fontenc}

\DeclareRobustCommand{\VAN}[3]{#2}
\let\VANthebibliography\thebibliography
\def\thebibliography{\DeclareRobustCommand{\VAN}[3]{##3}\VANthebibliography}

\usepackage{graphicx}	
\usepackage{amsmath}	
\DeclareMathAlphabet{\pazocal}{OMS}{zplm}{m}{n}

\title[The proper motion anomaly of HW Virginis]{New evidence about HW Vir's circumbinary planets from \textit{Hipparcos-Gaia} astrometry and a reanalysis of the eclipse timing variations using nested sampling}

\author[T A. Baycroft et al.]{
Thomas A. Baycroft,$^{1}$\thanks{E-mail: txb187@student.bham.ac.uk}
Amaury H.M.J. Triaud,$^{1}$
Pierre Kervella,$^{2}$
\\
$^{1}$School of Physics and Astronomy, University of Birmingham, Edgbaston, Birmingham B15 2TT, UK\\
$^{2}$LESIA, Observatoire de Paris, Universit\'e PSL, CNRS, Sorbonne Universit\'e, Universit\'e Paris Cit\'e, 5 place Jules Janssen, 92195 Meudon, France\\
}

\date{Accepted XXX. Received YYY; in original form ZZZ}

\pubyear{2023}

\begin{document}
\label{firstpage}
\pagerange{\pageref{firstpage}--\pageref{lastpage}}
\maketitle

\begin{abstract}
The post common-envelope eclipsing binary HW Virginis has had many circumbinary companions proposed based on eclipse timing variations. Each proposed solution has lacked in predictability and orbital stability, leaving the origin of the eclipse timing variations an active area of research. Leveraging the catalogue of \textit{Hipparcos} and \textit{Gaia} proper motion anomalies, we show there is slight evidence for a circumbinary companion orbiting HW Vir. We place an upper limit in mass for such a companion which excludes some previously claimed companions. We also apply this method to V471 Tauri and confirm the non-detection of a previously claimed brown dwarf. We adapt the {\tt kima} nested sampling code to analyse eclipse timing variations and re-analyse archival data on HW Vir, varying the order of the ephemeris that we fit for and the amount of the data that we use. Although signals are clearly present, we find two signals around 2500 and 4000 day periods that are not coherent between different \textit{chunks} of the data, so are likely to not be of planetary origin. We analyse the whole dataset and find the best solution to contain four signals. Of these four we argue the outermost is the most compatible with astrometry and thus the most likely to be of planetary nature. We posit the other three pseudo-periodic signals are caused by physical processes on the white dwarf. The eventual release of the full \textit{Gaia} epoch astrometry is a promising way to confirm whether circumbinary planets exist around HW Vir (and other similar systems), and explore white dwarf physics.
\end{abstract}

\begin{keywords}
 binaries:close -- binaries: eclipsing -- astrometry -- planets and satellites: detection -- stars: individual: HW Vir -- stars: subdwarfs.
\end{keywords}



\section{Introduction}

Although the majority of known exoplanets have been detected around single stars on the main sequence, planetary systems around post-main sequence stars and in binary star systems are known to exist. The first detected exoplanetary system was around a pulsar \citep{wolszczan_planetary_1992} and planets orbiting single white dwarfs are known to exist \citep[e.g.][]{bachelet_moa_2012,vanderburg_disintegrating_2015,vanderburg_giant_2020}. Many single white dwarf stars have been found to exhibit irregular transit-like and dimming events as well as having atmospheres polluted with heavy elements, both pointing to debris being accreted onto the star which could potentially have been scattered inwards by an invisible companion \citep{koester_frequency_2014,farihi_relentless_2022}. Planetary systems around main sequence binaries have been detected in transit by Kepler \citep[e.g.][]{doyle_kepler-16_2011} and \textit{TESS} \citep[e.g.][]{kostov_toi-1338_2020} and also in radial velocity \citep[e.g.][]{standing_radial-velocity_2023}. 

Planets are therefore known to orbit main-sequence binaries and are able to survive the evolution of a single star. There have been many claims of planets\footnote{Many of these have masses that would put them above the Deuterium burning limit and should be referred to as brown dwarfs, however for simplicity we refer to call them all "planets".} orbiting evolved binaries, but they are yet to be fully confirmed as planets. These candidate planets, orbiting post-common envelope binaries, are currently claimed based on periodic variations of the binary's mid-eclipse times. These variations can arise due to the light travel-time effect (LTTE) from the eclipsing binary orbiting the common center-of-mass between itself and the companion. These putative planets could be the counterparts of the detected main-sequence circumbinary planets that have lived through the evolution of their host binary \citep[e.g.][]{columba_statistics_2023}.

The existence of these planets has been debated \citep{mustill_main-sequence_2013}. In many cases the claimed planetary solutions fail to predict future eclipses \citep[e.g.][]{pulley_eclipse_2022}. One candidate, orbiting V471 Tauri \citep{beavers_v471_1986} was later followed up with direct imaging and was not detected with a high confidence \citep{hardy_first_2015}.

HW Virginis (HW Vir) is one of the most famous examples of post-common envelope binaries with claimed companions, first proposed by \citet{lee_sdbm_2009}. The system consists of a sdB primary of mass \(M_{\rm A} = 0.418\pm0.008\,\mathrm{M_{\odot}}\) and an M-dwarf secondary of mass \(M_{\rm B} = 0.128\pm0.004\,\mathrm{M_{\odot}}\) in a binary with orbital period \(P_{\rm bin} = 0.116719556\pm7.4\times10^{-9}\,\mathrm{days}\)\footnote{Parameters are taken from \citet{esmer_revisiting_2021}, these values are used for the rest of the analysis when the mass of the central binary is needed}. Eclipses have been precisely measured for over 30 years, with many conflicting solutions proposed \cite[e.g.][]{beuermann_quest_2012,esmer_revisiting_2021} with either one or two planets proposed as the cause of the eclipsing timing variations. One major issue is that none of the single-planet solutions fit the data satisfactorily, but none of the better-fitting two-planet solutions appear to be dynamically stable \citep{brown-sevilla_new_2021,mai_eclipse_2022}. Another issue, as mentioned above, is that all of the proposed solutions very quickly diverge from the data subsequently collected.

Non-planetary explanations have been suggested which can produce eclipse timing variations in short-period binaries such as HW Vir. The period (or apparent period) of the binary could be affected by apsidal precession if it is eccentric, and magnetic braking \citep{rappaport_new_1983} or emission of gravitational waves \citep{paczynski_gravitational_1967} could cause the orbit to shrink due to angular momentum loss. Other magnetic effects have also been proposed, such as the Applegate mechanism \citep{applegate_mechanism_1992}, or a more recent mechanism, requiring less energy suggested by \citet{lanza_internal_2020}. However in most cases these are insufficient to fully explain the shape or the amplitude of the observed modulations in eclipse time. 

Many of these candidate planets will need to be confirmed/rejected through other methods. One example of this happening is with V471 Tau. This system consists of a WD primary of mass \(M_{\rm A} = 0.797\pm0.016\,\mathrm{M_{\odot}}\) and a K-Dwarf secondary of mass \(M_{\rm B} = 0.864\pm0.029\,\mathrm{M_{\odot}}\) in a binary with orbital period \(P_{\rm bin} = 0.5211834194\pm7.2\times10^{-9}\,\mathrm{days}\) \citep{muirhead_revised_2022}. This system shows periodic variations of the mid-eclipse times, which have been used to suggest an orbiting brown dwarf \citep{beavers_v471_1986,guinan_best_2001}. The system has since been directly imaged with SPHERE, and these observations resulted in a non-detection \citep{hardy_first_2015}, thus rejecting the claimed brown dwarf.

Planets around ultra-short period evolved binaries such as these may also eventually be detectable in gravitational waves, for example by the Laser Interferometer Space Antenna, LISA \citep{danielski_circumbinary_2019}. LISA will however only be sensitive to binaries of shorter orbital period than HW Vir.

Another possibility for confirming or rejecting post-common envelope circumbinary planets is precise astrometry. The space telescope \textit{Gaia} \citep{gaia_collaboration_gaia_2021} is performing a precise astrometric survey of the whole sky which will have a baseline of about 10 years. \textit{Gaia}'s astrometric solution will be able to investigate some of these systems without relying on any eclipse timing data \citep{sahlmann_gaias_2015}. However, individual astrometric measurements are expected to released around late 2025. In the meantime, we can only rely on the proper motion anomaly method \citep{kervella_stellar_2019}. Before \textit{Gaia}, \textit{Hipparcos} \citep{esa_hipparcos_1997} also performed an astrometric survey, but of a much smaller sample of stars, at a lower precision. HW Vir is within that sample, as is V471 Tau. The proper motion anomaly method combines positions and proper motions of a star from \textit{Hipparcos} and one of \textit{Gaia}'s recent data releases to estimate the effect of an orbiting of an orbiting companion on the proper motion of the star. This method has been applied to single stars,  and combined with other techniques such as radial velocity, has led to the detection and characterisation of several planetary companions \citep[e.g.][]{mesa_constraining_2022,rickman_precise_2022}.

In this paper we present a new piece of astrometric information, in the form of the proper motion anomaly, to the puzzle that is HW Vir, and we perform a new fit of the eclipse timing data utilising nested sampling and analysing different \textit{chunks} of data separately. We report on the lack of consistency of signals in the eclipse times and present the model that we find to fit best to the whole dataset, providing a suggestion of which signal is favoured to be planetary since not all the detected signals can be. 

The paper is set out as follows. We describe the proper motion anomaly method, and apply it to both V471 Tau and  HW Vir in Section \ref{sec:ast}. In Section \ref{sec:fit}, the use of {\tt kima} to fit eclipse times is described. Section \ref{sec:ETVs} details the eclipse timing data used, and the results from the analysis of the data. We discuss the results and implications and conclude in Section \ref{sec:disconc}. 

\section{Astrometry: using the proper motion anomaly method}\label{sec:ast}

We firstly explain the method of the astrometric proper motion anomaly, and secondly apply this to both  V471 Tau and HW Vir. We compare the results from the astrometric proper motion anomaly to some of the previously proposed planetary solutions.

\subsection{How does the proper motion anomaly work?}

The proper motion anomaly analysis method is described in detail in \citet{kervella_stellar_2019}.
Using positional measurements from \textit{Gaia} and \textit{Hipparcos}, we determine the long-term, mean proper motion vector of the system $\mu_\mathrm{HG}$ by dividing the observed change in position by the time baseline $\delta t_\mathrm{HG}$ between the two measurements (that is, 24.75 years between \textit{Hipparcos} and \textit{Gaia} DR3). 
For the nearest stars, second order effects must be taken into account in this computation, but they are negligible for the systems discussed in the present paper.

\begin{figure}
    \centering
    \includegraphics[width=\columnwidth]{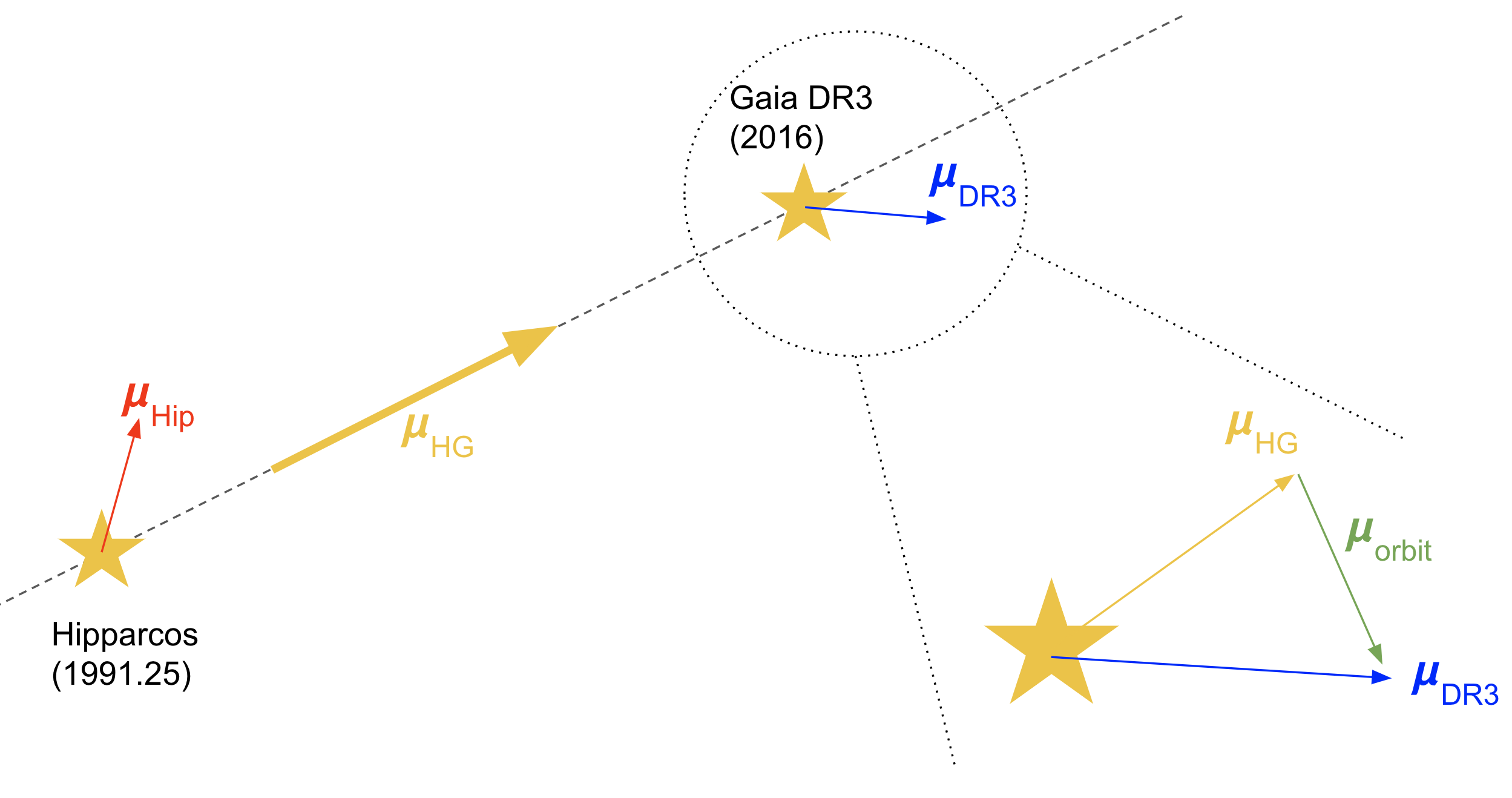}
    \caption{Diagram of the proper motion anomaly method.}
    \label{fig:pma_diagram}
\end{figure}

Thanks to the long time baseline, and assuming that the orbital period of the companion is significantly shorter than $\delta t_\mathrm{HG}$, $\mu_\mathrm{HG}$ essentially traces the proper motion of the center of mass of the system.
Separately, the short-term proper motion measurements $\mu_{\rm Hip}$ and $\mu_\mathrm{DR3}$ (obtained respectively by \textit{Hipparcos} and \textit{Gaia}) trace the vector sum of 1) the linear proper motion of the center of mass and 2) the orbital motion $\mu_\mathrm{orbit}$ of the photocentre of the system around the center of mass.
Figure~\ref{fig:pma_diagram} visually presents the different vector quantities considered in these computations.

When considering a planetary companion, the photocentre is located very close to the geometrical center of the star.
In this configuration, subtracting the long-term proper motion $\mu_\mathrm{HG}$ from the \textit{Gaia} DR3 short-term proper motion $\mu_\mathrm{DR3}$ gives access to the proper motion anomaly of the star $\Delta \mu$, that traces the orbital motion of the star around the center of mass of the system.
The quantity $\Delta \mu$ can be scaled to a linear tangential velocity anomaly $v_\mathrm{tan}$ using the parallax. This is the two-dimensional counterpart of the radial velocity that is traditionally employed to detect exoplanets.

Corrective terms must be considered to interpret the measured proper motion anomaly in terms of companion properties. Firstly, the $\mu_\mathrm{orbit}$ quantity is an average quantity over the \textit{Gaia} integration window, that has a duration of $\delta t_\mathrm{DR3} = 34$\,months.
This averaging implies that the measured proper motion anomaly will be smeared, reducing the sensitivity in terms of companion mass.  This loss of sensitivity is particularly strong for orbital periods shorter than \(\delta t_\mathrm{DR3}\)
Secondly, the time baseline $\delta t_\mathrm{HG}$ between \textit{Hipparcos} and \textit{Gaia} DR3 results in the subtraction of part of the proper motion signature of very long period companions ($P>3\times \delta t_\mathrm{HG}$) during the computation of the $\mu_\mathrm{HG}$ quantity. This effect induces a loss of sensitivity to such very long period companions.
These two effects determine the companion mass sensitivity function of the proper motion anomaly method.

We now derive the equation for the tangential velocity caused by a companion if measured instantaneously. This equation can then be combined with the sensitivity function calculated numerically. The proper motion is usually divided into its components in right-ascension (ra) and declination (dec). 

\begin{equation}
    \bf{\mu} = \mu_{\rm ra}\bf{e}_{\rm ra} + \mu_{\rm dec}\bf{e}_{\rm dec},
\end{equation}
with \(\bf{e}_{\rm ra}\) and \(\bf{e}_{\rm dec}\) the basis vectors in ra and dec. We then subtract the long-term HG proper motion from the \textit{Gaia} DR3 proper motion and take the magnitude of this vector to get the tangential velocity anomaly.
\begin{align}
    \bf{\Delta \mu} &= (\mu_{\rm DR3, ra}-\mu_{\rm HG, ra})\bf{e}_{\rm ra} + (\mu_{\rm DR3, dec}-\mu_{\rm HG, dec})\bf{e}_{\rm dec},\\
    v_{\rm tan} &= \frac{1}{\varpi}\sqrt{(\mu_{\rm DR3, ra}-\mu_{\rm HG, ra})^2 + (\mu_{\rm DR3, dec}-\mu_{\rm HG, dec})^2},
\end{align}
where \(\varpi\) is the parallax.
Now given an inner mass \(M\) and an outer mass \(m\) the relative orbital velocity is
\begin{equation}
    V = \sqrt{G(M+m)}\left(\frac{2}{r} - \frac{1}{a}\right)^{1/2},
\end{equation}
where G is the gravitational constant, \(a\) the semi-major axis of the relative  orbit, and \(r\) the relative orbital distance at the measured time. The distance is given by
\begin{equation}
    r = \frac{a(1-e^2)}{1+e\cos f},
\end{equation}
with \(e\) and \(f\) the eccentricity and true anomaly of the orbit at the measured time (they are the same for the relative orbit or the orbit of one of the components). Combining these two equations and using that the velocity of the inner body (i.e. the luminous one) relates to the outer velocity by
\begin{equation}
    v_0 = \frac{m}{M+m}V,
\end{equation}
gives us:
\begin{align}
    v_{\rm tan} &= \sqrt{\frac{Gm^2}{a(M+m)}}\left(\frac{2(1+e\cos f)}{(1-e^2)} - 1\right)^{1/2}\\
    v_{\rm tan} &= \sqrt{\frac{Gm^3}{a_1(M+m)^2}}\left(\frac{2(1+e_1\cos f_1)}{(1-e_1^2)} - 1\right)^{1/2},
\end{align}
where \(a_1\) is the semi-major axis of the outer orbit which relates to that of the relative orbit by \(a_1 = \frac{m}{M+m}a\).

This derivation is valid for an instantaneous measurement of \(v_{\rm tan}\), which would correspond to an instantaneous measurement of \(\mu_{\rm DR3}\) and an infinitely long baseline for \(\mu_{\rm HG}\). This is, of course, not the case. As described above we must also include a sensitivity function. This function has been numerically calculated by \citet{kervella_stellar_2019}. This leads to the sensitivity curve for the proper motion anomaly at different periods which is used in the following section. These curves give the areas of Period-Mass space that are consistent with a measured proper motion anomaly, under the assumption of a circular orbit.

\subsection{Applying the proper motion anomaly to V471 Tau and HW Vir}
Calculating the long-term proper motion between \textit{Hipparcos} and \textit{Gaia} DR3 has been done and combined into a catalogue by both \citet{kervella_stellar_2022} and \citet{brandt_hipparcos-gaia_2021}, using different combinations of the main \textit{Hipparcos} reductions. The values for the tangential velocity for HW Vir and V471 Tau are shown in Table \ref{tab:PMa_values}. We note that the values are in good agreement between both catalogues and choose arbitrarily to use the \citet{kervella_stellar_2022} value from now on.

\subsubsection{V471 Tau}

\begin{figure}
    \centering
    \includegraphics[width=\columnwidth]{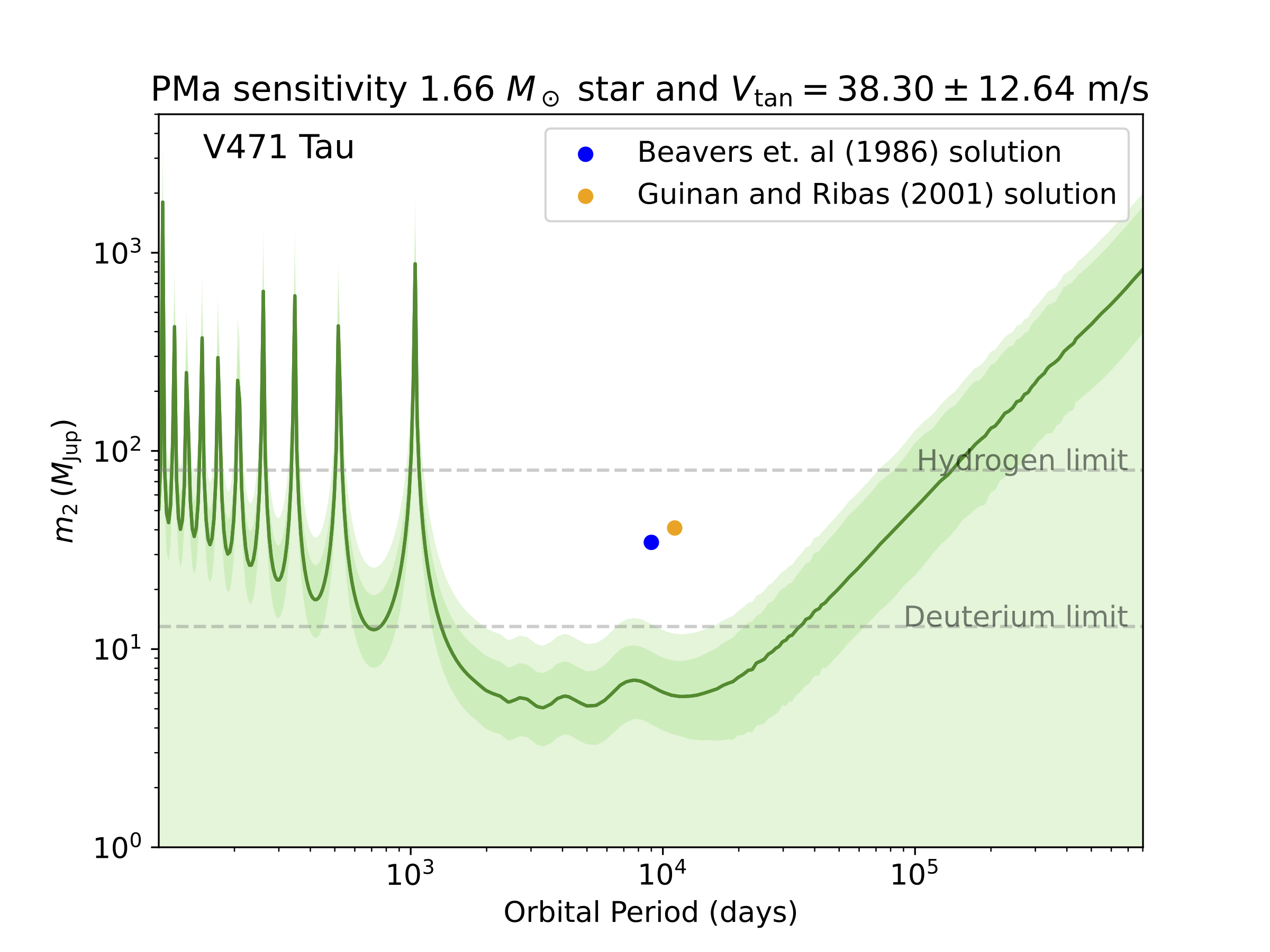}
    \caption{Sensitivity curve for proper motion anomaly applied to V471 Tau. Green shows the mean, 1-\(\sigma\) region and 3-\(\sigma\) region of parameter space that could correspond to an orbiting body giving rise to the proper motion anomaly. The coloured dots show locations of claimed solutions by 2 previous papers. The dashed lines show the locations of the hydrogen and deuterium fusing limits.}
    \label{fig:V471}
\end{figure}

For V471 Tau, we have a proper motion anomaly between 2-\(\sigma\) and 3-\(\sigma\). Figure \ref{fig:V471} shows the sensitivity curve of the proper motion anomaly method associated with the value for V471 Tau from \citet{kervella_stellar_2022}. The dark green line shows the curve on which a body needs to lie to produce the observed tangential velocity value, the darker and lighter shaded regions show the \(1\sigma\) and \(3\sigma\) regions around that line. The spikes towards shorter orbital periods are the result of the sensitivity function as described above. The slope at longer orbital periods is produced when only small fraction of an orbital arc is covered and hence the efficiency function is small. In between, is a region of highest sensitivity. This provides an upper bound that is far below the mass of the proposed solutions by \citet{beavers_v471_1986} and \citet{guinan_best_2001}. This re-affirms the conclusion of \citet{hardy_first_2015} which did not find evidence of the proposed brown dwarf, and confirms that the variations in the mid-eclipse times must be coming from some other source.

We numerically estimate the proper motion anomaly that would be caused by the binary. For a given set of parameters (\(M_0,\,M_1,\,P\)) we perform a bisection algorithm suggesting values for \(v_{\rm tan}\), and comparing the value of \(M_1\) obtained (given \(M_0\) and \(P\)) to the given value, until the masses agree to \(0.001 {\rm M_{jup}}\). We repeat this for 1000 realisations of the binary parameters to then obtain the median and 1\(\sigma\) values of \(32^{+13}_{-21}\) \(\mathrm{m\,s^{-1}}\). This is entirely consistent with the tentative signal that is seen. The proper motion anomaly is sensing the smeared binary motion and so does not suggest an orbiting companion.

\subsubsection{HW Vir}
For HW Vir, the tangential velocity is distinct from zero at around \(2\sigma\) confidence in both catalogues. We cannot therefore conclude from the proper motion anomaly that there is definitely an orbiting body, but this {\it Gaia}-{\it Hipparcos} combined measurement brings new evidence that suggests such a body is more likely to exist than not.

First, we validate that this tentative proper motion anomaly is not caused by the smeared orbital motion of binary. In the same way as for V471 Tau, we numerically estimate the proper motion anomaly that would be induced by the binary and obtain the median and \(1\sigma\) values of \(2.52^{+0.80}_{-1.60} \,\mathrm{m\,s^{-1}}\). The excess tangential velocity is therefore not caused by the HW Vir binary.

The top panel of Figure \ref{fig:periodogram} shows the sensitivity curve of the proper motion anomaly method associated with HW Vir. The curve has the same shape as in Figure \ref{fig:V471} (since all the spikes are primarily related with the {\it Gaia} 34-month observing window) but is zoomed in on the area of best sensitivity. We overplot the locations of the orbiting bodies proposed by three previous studies \citep{beuermann_quest_2012,brown-sevilla_new_2021,esmer_revisiting_2021}. We note that four of the proposed solutions include one orbiting body above the \(3\sigma\) line. These solutions are disfavoured\footnote{They may still be possible in reality, if we have a very eccentric orbit for the companion, and we observe it close to apastron (where the motion is slower)} by the observed proper motion anomaly which is too weak to have been produced by these putative objects. This plot also shows the locations of the four components from our best-fitting model\footnote{We do not claim that all four of the signals are indeed planets}, which we describe in Section \ref{sec:results_full}.
\begin{table}
    \centering
    \begin{tabular}{l|c|c|r}
    \hline
    & HW Vir & V471 Tau & \\
    \hline
    \citet{kervella_stellar_2022}  & \(214\pm111\)  & \(38\pm13\) & \(\mathrm{m\,s^{-1}}\) \\
    \citet{brandt_hipparcos-gaia_2021} &  \(226\pm111\)  & \(28\pm13\)& \(\mathrm{m\,s^{-1}}\) \\
    \hline
    \end{tabular}
    \caption{Tangential velocities from the proper motion anomaly between \textit{Gaia} DR3 and the \textit{Hipparcos}-\textit{Gaia} long-term vector, for HW Vir and V471 Tau. Values are reported using both the \citet{kervella_stellar_2022} and \citet{brandt_hipparcos-gaia_2021} catalogues of accelerations.}
    \label{tab:PMa_values}
\end{table}

This tentative proper motion anomaly is an extra piece of information about the HW Vir system which provides astrometric evidence that there may be an orbiting circumbinary companion.

The catalogues of accelerations mentioned above rely on \textit{Gaia} EDR3 \citep{gaia_collaboration_gaia_2021}. The same analysis was done earlier using \textit{Gaia} DR2 \citep{gaia_collaboration_gaia_2018}, and the value from \citet{kervella_stellar_2019} for HW Vir is \(309\pm200\). From this we infer that the astrometric signal is getting more confident as more \textit{Gaia} data becomes available. This implies that if there is indeed a signal there, it should be detectable from future \textit{Gaia} data releases.

\section{Fitting eclipse timing variations with {\tt kima}}\label{sec:fit}

Whilst verifying whether proposed solutions for the HW Vir systems were compatible with the proper motion anomaly, we also decided to re-analyse the eclipse times of HW Vir with a nested sampler, which we believe has not been attempted yet. Most of the literature uses $\chi^2$ maps or reduced $\chi^2_\nu$ to make inferences about the number of signals present in the data, but none have conducted a Bayesian model comparison in this way yet.

Amongst Bayesian methods, nested sampling has the advantage to let some key parameters free that are usually fixed in other types of analyses. In our case, the number of orbiting planets, $N_{\rm p}$ is a free parameter which allows for a robust model comparison, based on a ratio of Bayesian evidence. All planetary signals are adjusted at once and models with 0, 1, 2... planets are constantly compared to one another.

{\tt kima} is an orbital fitting algorithm originally designed for application to radial velocities \citep{faria_kima_2018}. We adapt it to fit mid-eclipse times instead, to then apply it to HW Vir. {\tt kima} leverages nested sampling using DNEST4 \citep{brewer_dnest4_2018} to explore parameter space and calculate the likelihood of proposed samples. Using the trans-dimensional sampling in {\tt kima} the number of Keplerian signals, \(N_{\rm p}\), being fit is a free parameter as described above. This allows a comparison of the different numbers of signals present in the data with a Bayes Factor. The Bayes Factor for a \(N_{\rm p}=n\) model\footnote{\(N_{\rm p}\) meaning number of planets in the model} compared to a \(N_{\rm p}=n-1\) model is the ratio of the evidence \(Z\) for each model. 

The evidence is the primary output of nested sampling and is the integral of the likelihood over the prior mass. In nested sampling this integral is calculated as a weighted sum, with the weights being associated to the change in prior mass between consecutive samples \citep{skilling_nested_2006}. In this case the evidence for an \(N_{\rm p}=n\) model is the sum of the \textit{weights} of all the samples with \(n\) planets, and then the Bayes Factor is \(BF = \frac{Z_n}{Z_{n-1}}\).

We use a detection threshold of 150 as recommended by \citet{kass_bayes_1995}. A Bayes Factor larger than 150 is taken as very strong evidence for the more complex model over the less complex model (and roughly equivalent to a p-value of 0.001). It is common that a nested sampler finds the sum of the \textit{weights} of all the samples is highest for the highest \(N_{\rm p}\) explored by the sampler  \citep{faria_kima_2018,standing_bebop_2022}. However, so long as the ratio is not \(>150\) those most complex solutions, whilst providing a better fit to the data, do not contain enough statistical evidence to warrant the extra number of parameters.

As a by-product of the nested sampling to calculate the evidences, we can obtain posterior samples for the various parameters from {\tt kima}. These allow us to perform parameter estimation on any detected signals, assuming a light travel-time effect (LTTE) model with a companion on a Keplerian orbit. {\tt kima} has already been used to detect and test the detectability of circumbinary planets with radial-velocities \citep{triaud_bebop_2022,standing_bebop_2022,standing_radial-velocity_2023}. We redirect the reader to these publications for more thorough explanations of how the model comparison works.

We fit the eclipse times in {\tt kima} with a number of Keplerian signals as well as an ephemeris function. We allow the ability to fit for one of a linear, quadratic or cubic ephemeris. These are shown in equation \ref{eq:etv_fit} below:

\begin{equation}
    \label{eq:etv_fit}
    T(E) = T_0 + P_0E + \frac{1}{2}\dot{P}_0P_0E^2 + \frac{1}{6}\ddot{P}_0P_0^2E^3 + \sum_i \tau_i(E),
\end{equation}
where \(E\) is the epoch of an eclipse (i.e. the number of the eclipse with the first eclipse being 0), \(T(E)\) is the time of that eclipse in our model, \(T_0\) is the reference time (time at epoch 0 here), \(P_0\) is the period of the eclipsing binary at the reference time, \(\dot{P}_0\) and \(\ddot{P}_0\) are the first and second time-derivatives of the binary period (at the reference time), and \(\tau_i\) is the time-delay due to the LTTE of an orbiting body. The middle three terms are a Taylor series and if we ignore the terms of order \(\geq E^2\) we are using a linear ephemeris, if we ignore terms of order \(\geq E^3\) we are using a quadratic ephemeris, and using all the terms above is a cubic ephemeris. The functional form for the LTTE due to an orbiting body is as in \citet{irwin_determination_1952}:

\begin{equation}
    \label{eq:tau_pl}
    \tau(t) = \frac{K}{\sqrt{1 - e^2\cos^2\omega}}\left(\frac{1-e^2}{1+e\cos\nu(t)}\sin(\nu(t)+\omega) + e\sin\omega\right),
\end{equation}
where \(e\) and \(\omega\) are the eccentricity and argument of periastron of the orbiting body, \(\nu(t)\) its true anomaly at time \(t\) and \(K\) the semi-amplitude of the signal:
\begin{equation}
    K = \frac{m \sin i}{c(M+m)}\left(\frac{G(M+m)}{4\pi^2}\right)^{1/3}P^{2/3},
\end{equation}
where \(m\) and \(P\) are the mass and orbital period of the orbiting body, \(M\) the total mass of the eclipsing binary, i the inclination to the line of sight of the planetary orbit, and \(c\) and \(G\) the speed of light and gravitational constant.

In equation \ref{eq:tau_pl}, \(\tau\) and \(\nu\) are functions of \(t\) (time). The orbital period of an orbiting body is much greater than the difference in time from all other terms, so we use \(t\approx P_0E\) as a first order approximation.

\section{Fitting Eclipse times}\label{sec:ETVs}

In this section we first detail from where the eclipse timing data is obtained, and then present the results from the analysis using {\tt kima}.

\subsection{Data for HW Vir}\label{sec:data}

We use archival data for HW Vir eclipse times (considering only the primary eclipse). We use the data from \citet{brown-sevilla_new_2021}, which collated data from \citet{kilkenny_period_1994}, \citet{lee_sdbm_2009}, and \citet{beuermann_quest_2012}, as well as their own data. To this we add the data from \citet{baran_pulsations_2018}, \citet{esmer_revisiting_2021}, and \citet{mai_eclipse_2022}. Of the data reported in \citet{baran_pulsations_2018}, we find that the data taken with SAAO have a small offset of \(\sim 80\) sec from the rest of the datasets (including the other data reported in the same publication). Since there is still good coverage without this, we exclude these data from our analysis.

We perform the analysis on the whole dataset, but also divide it into smaller \textit{chunks} to assess how consistent any signals that appear are. This way we can assess if, although the overall model doesn't have good predictive power, a subset of the signals might be predictably and consistently present. We divide the dataset into \textit{chunks} of approximately 1/3 and 2/3 the length of the whole dataset, with epochs as shown in Table \ref{tab:data_division}. The \textit{chunk} "tier3", is extended back an extra 10\,000 epochs and overlaps with "tier2". The different \textit{chunks} are also visualised alongside the data in the top panel of Figure \ref{fig:corners}.


\begin{table}
    \centering
    \caption{Division of the eclipse time data into \textit{chunks}, the names of the \textit{chunks} are specified as well as the corresponding epoch ranges included in each one. The data is effectively partitioned into thirds.}
    \begin{tabular}{l|r}
        \hline
        \textit{chunk} & epoch range \\
        \hline
        tier1 & \(0\leq E < 40000\) \\
        tier2 & \(40000 \leq E < 80000\)\\
        tier3 & \(70000 \leq E \)\\
        tier1-2 & \(0\leq E < 80000\) \\
        tier2-3 & \(40000\leq E \) \\
        tier1-3 & \(E \leq 40000 \) or \(80000\leq E \) \\
        full & \(0 \leq E \) \\
        \hline
    \end{tabular}
    \label{tab:data_division}
\end{table}

\begin{figure*}
    \includegraphics[width=2\columnwidth]{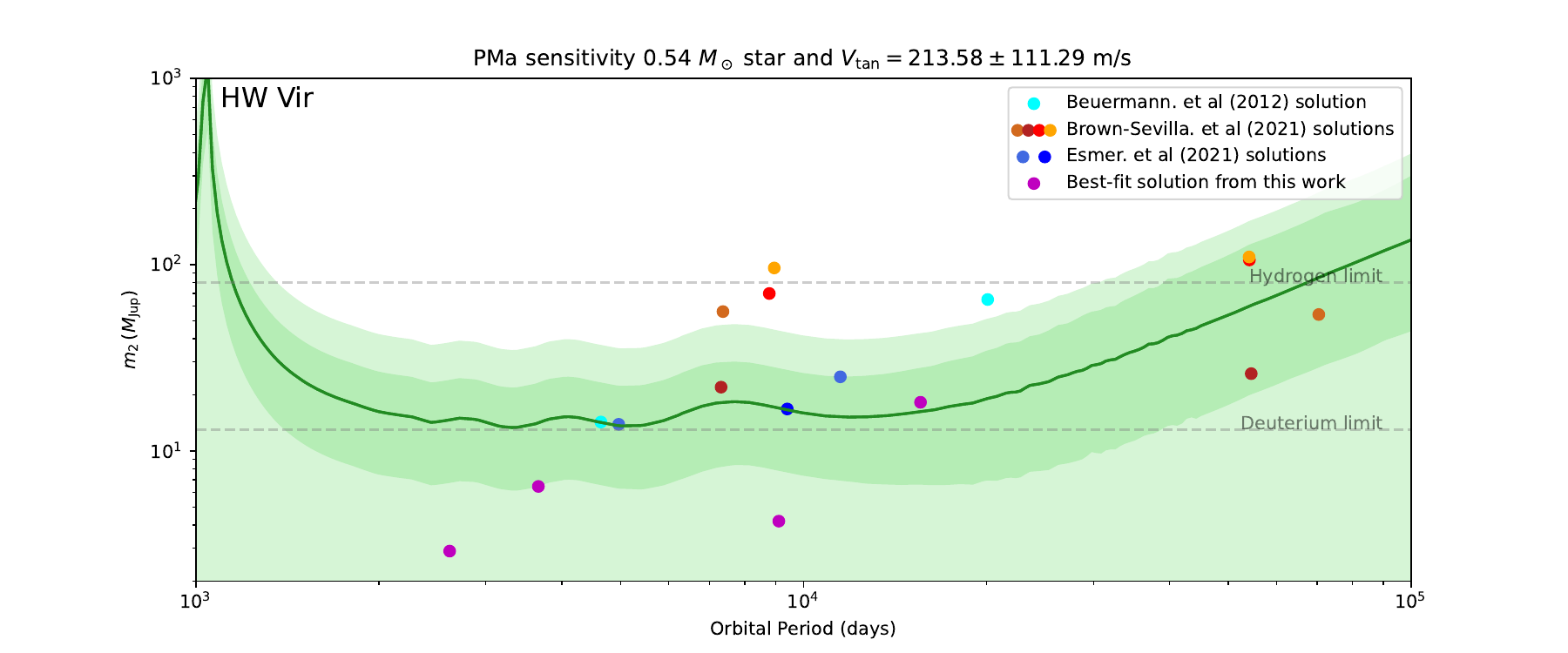}\\
    \includegraphics[width=2\columnwidth]{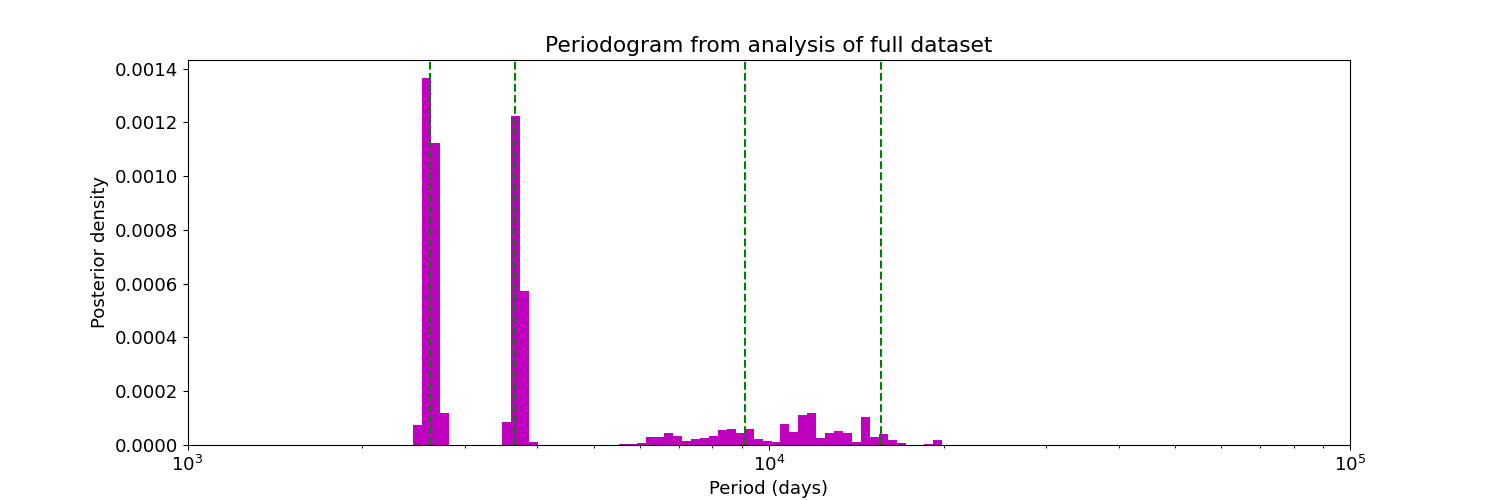}\\
    \caption{Top: Sensitivity curve for proper motion anomaly applied to HW Vir. Green shows the mean, 1-\(\sigma\) region and 3-\(\sigma\) region of parameter space that could correspond to an orbiting body giving rise to the proper motion anomaly. The coloured dots show locations of claimed solutions by 3 previous analyses as well as the best-fitting solution from this work. The dashed lines show the locations of the hydrogen and deuterium fusing limits. Middle: posterior density histogram of the periods of planets suggested in all Np=4 posterior samples from the analysis of the full dataset. Dashed lines show the locations of the best fitting 4-planet solution.}
    \label{fig:periodogram}
\end{figure*}

\subsection{Results from eclipse timing variation fits}\label{sec:results}

In this section we present the results from a reanalysis of the mid-eclipse time data, analysed using {\tt kima}. The nested sampling implemented requires a prior distribution for each parameter, these are detailed in Table \ref{tab:priors}. The Kumaraswamy distribution \citep{kumaraswamy_generalized_1980} approximates the beta distribution, and the shape parameters as shown in Table \ref{tab:priors} are those that \citet{kipping_parametrizing_2013} argues best represent the distribution of exoplanetary eccentricities, based on exoplanets detected with the radial velocity method. The analysis is performed with each of a linear, quadratic, and cubic ephemeris for each \textit{chunk} as well as for the whole dataset.

\begin{table}
    \centering
    \caption{Prior distributions for the nested sampling analysis. \(\pazocal{N,\,LU,\,MLU,\,K}\) refer to Normal, log-Uniform, Modified log-Uniform (with a knee and an upper limit), and Kumaraswamy distributions, each taking two parameters. *This prior for Np is used in all cases except the analysis of the full dataset where instead the prior used is \(\pazocal{U}(0,6)\).}
    \begin{tabular}{c|c|c}
    \hline
    Parameter & Unit & Prior distribution\\
    \hline
    ephemeris parameters\\
    \(P_0\) & day & \(\pazocal{N}(0.11672,0.00001)\)\\
    \(P_{t,0}\)& & \(\pazocal{N}(0,0.00001)\)\\
    \(P_{tt,0}\) & day\(^{-1}\) &\(\pazocal{N}(0,0.0000001)\) \\
    \hline
    planet parameters\\
    \(P\) & day & \(\pazocal{LU}(500,20000)\)\\
    \(K\) & s &\(\pazocal{MLU}(0.1,10000\)\\
    \(e\) & & \(\pazocal{K}(0.867,3.03)\) \\
    \(\omega\) & rad & \(\pazocal{U}(0,2\pi)\) \\
    \(\phi_0\) & rad & \(\pazocal{U}(0,2\pi)\)\\
    \hline
    other parameters\\
    \(N_{\rm P}\) & & \(\pazocal{U}(0,3)\)* \\
    \(\sigma_{\rm jit}\) & s & \(\pazocal{MLU}(0.01,1000)\)\\
    \hline
    \end{tabular}
    \label{tab:priors}
\end{table}

\subsubsection{Results from analysing different \textit{chunks}}\label{sec:results_chunked}

A LTTE signal due to an orbiting companion should to be coherent in time. The analysis of different \textit{chunks}, using a Keplerian prescription, would therefore be expected to lead to posteriors that are consistent across \textit{chunks}. Our analysis of the different \textit{chunks} shows a lack of consistency and therefore casts doubts on the ETV signals being solely due to an orbiting companion (or more).

Throughout the analysis of the different \textit{chunks} of data, recurring signals are seen around two periods: 4000 days and 2500 days. Longer period (and higher amplitude) signals do exist in many of the \textit{chunks}, however they are far from consistent. 

\begin{figure*}
    \centering
    \includegraphics[width=2\columnwidth]{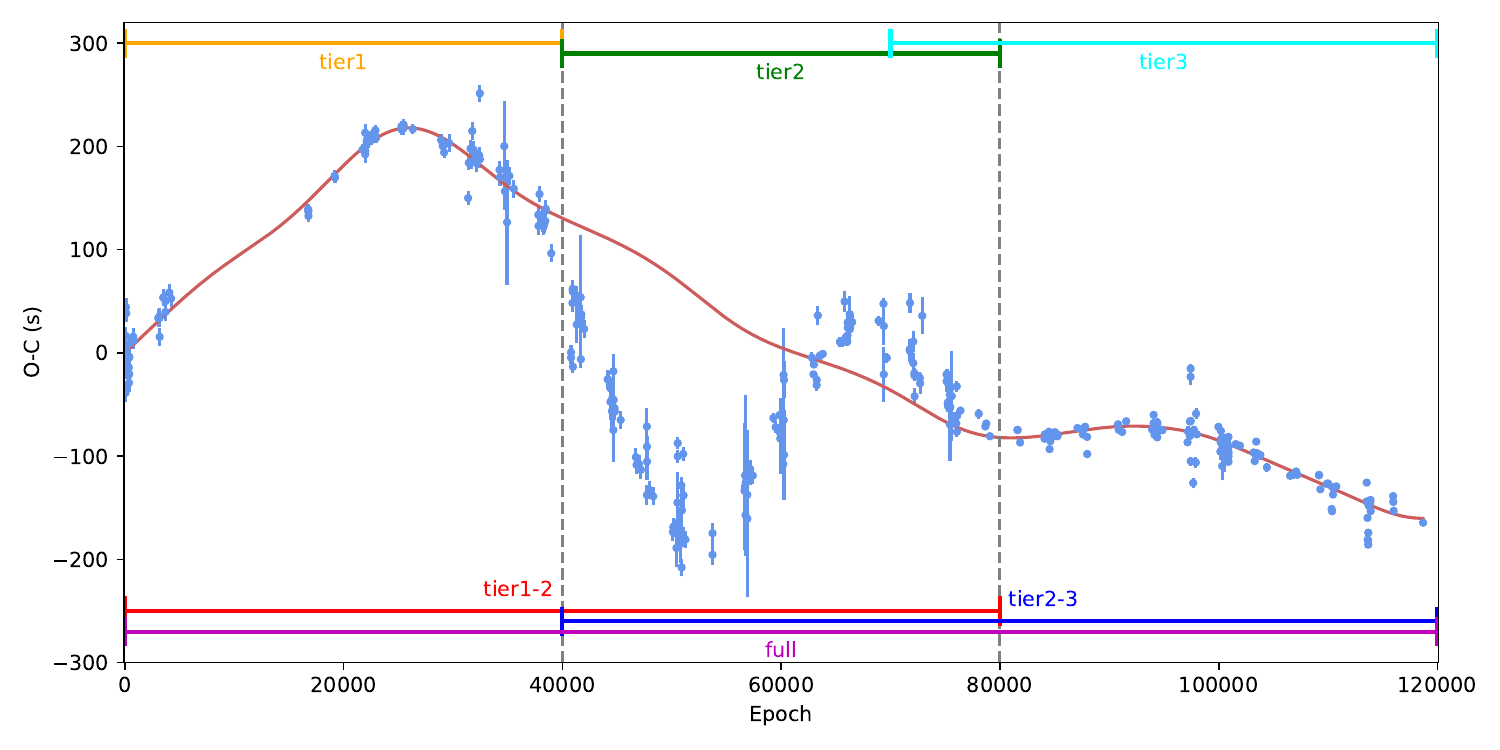}\\
    \includegraphics[width=1\columnwidth]{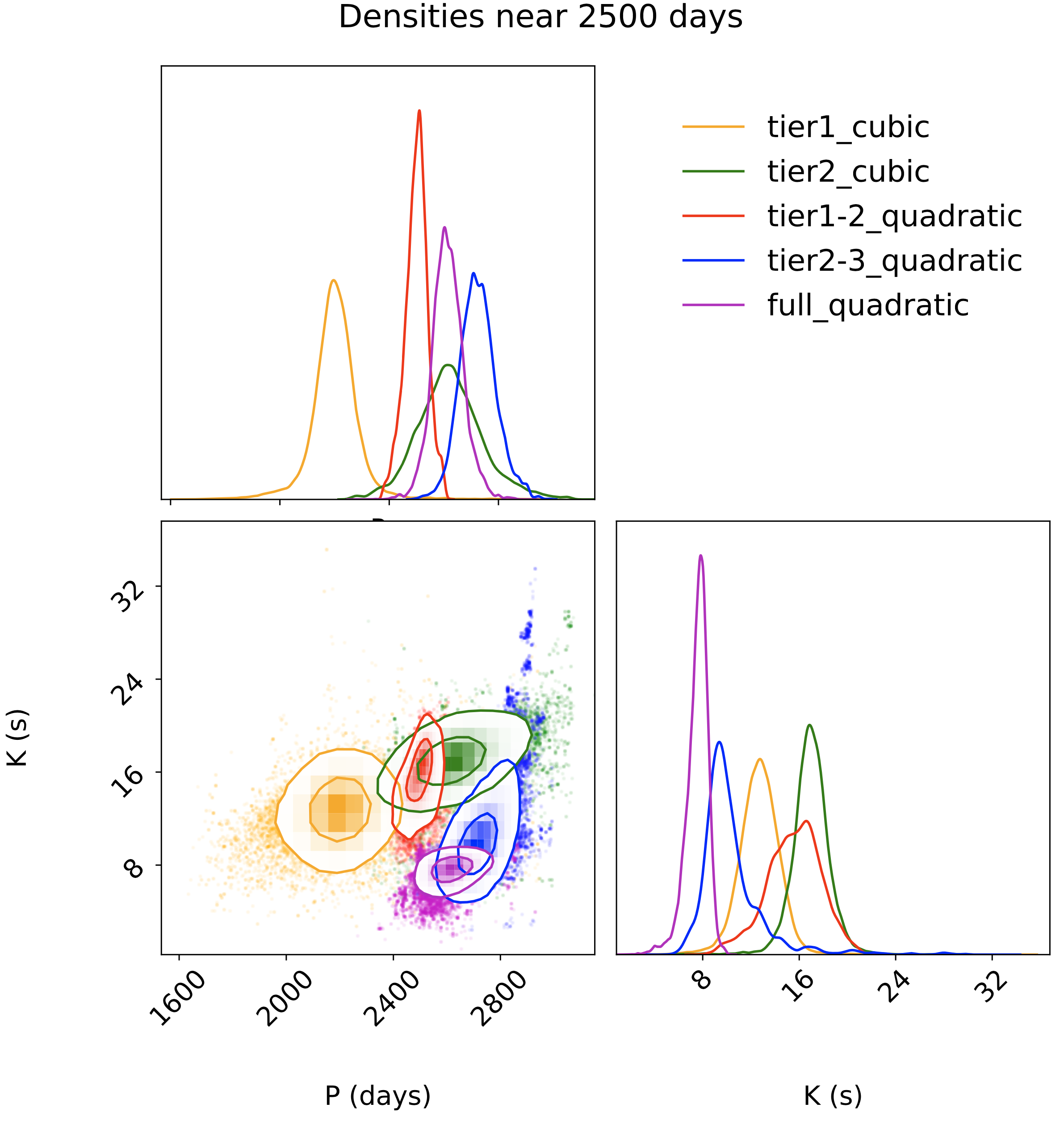}
    \includegraphics[width=1\columnwidth]{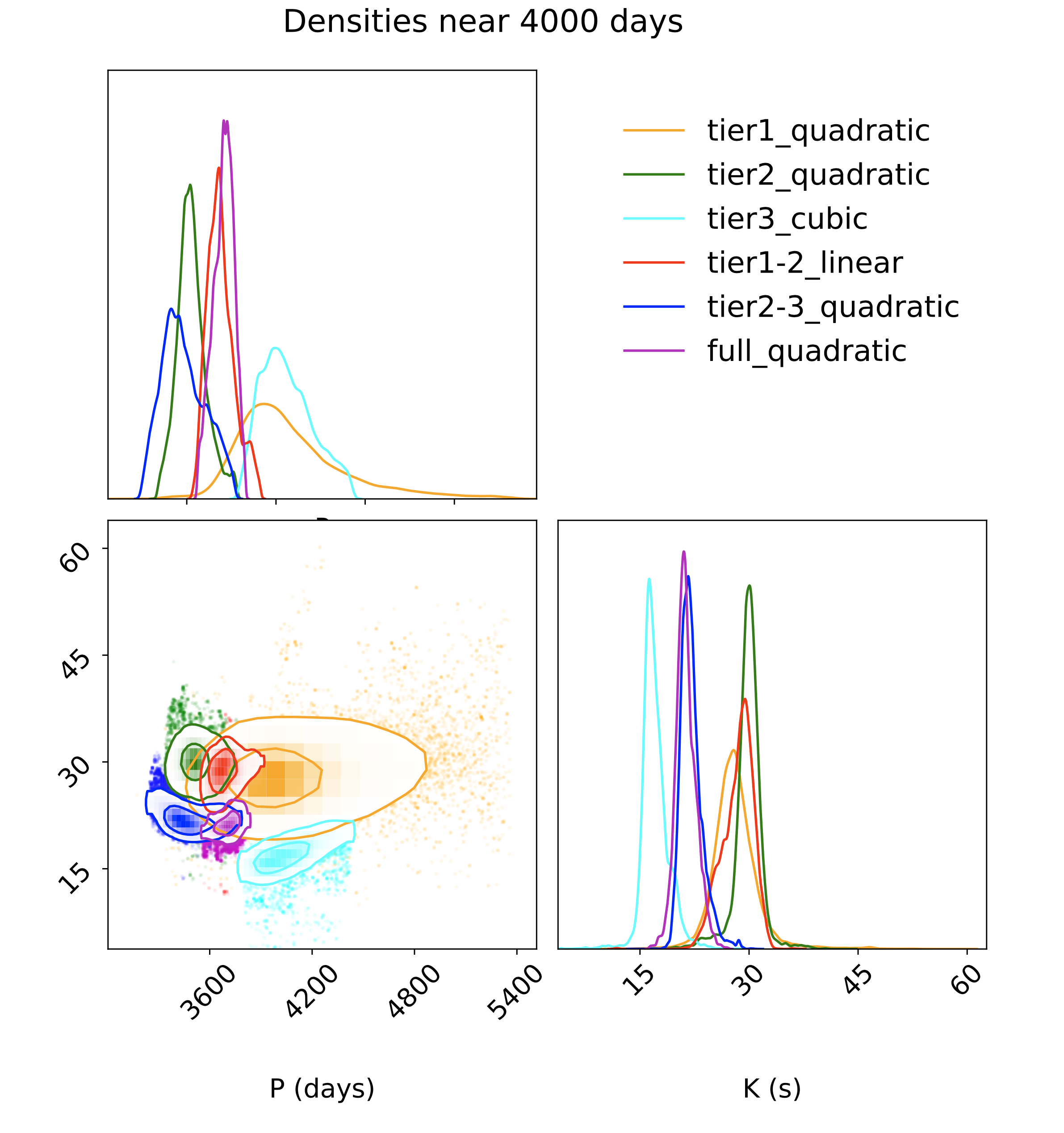}
    \caption{Top: Best fitting solution from the analysis of \textit{chunk} tier1-3, which is all the data except that which is between the dashed grey lines. The best-fitting model is shown in red, the full dataset (including the middle \textit{chunk} that was not included in the fit) is shown in blue. Highlighted in colour are the names and spans of data covered by the other \textit{chunks}. The x-axis is the Epoch \(E\) as in equation \ref{eq:etv_fit}. Bottom: Posterior density smoothed corner plots showing the period and semi-amplitude of signals found around 2500 days (left) and 4000 days (right). The different colours correspond to analysis of different \textit{chunks} of the data being analysed.}
    \label{fig:corners}
\end{figure*}

To assess the consistency of the signals at the recurring periods, the clustering algorithm HDBSCAN \citep{mcinnes_hdbscan_2017} is used to identify clusters in the P-K plane from the {\tt kima} posterior samples. The clusters are then visually associated with one of the two recurring periods or not. The lower panels of Figure \ref{fig:corners} show the clusters of posterior density around 2500 and around 4000 days from runs of {\tt kima} on different \textit{chunks} of data\footnote{tier3 did not show signature of a detectable signal around 2500 days so only 5 \textit{chunks} are shown}. These are shown as corner plots \citep{foreman-mackey_cornerpy_2016} between the period \(P\) and semi-amplitude \(K\). While there is a cluster of posterior density around this period in each\footnote{tier3 notwithstanding} of these runs, the periods and amplitudes of the signals vary between the runs.

The lack of consistency of these signals points to them not being due to a Keplerian LTTE orbit. These signals may then have a non-periodic or quasi-periodic source. If this is the case, then attempting to fit them with strictly periodic Keplerian signals is unideal. This is exemplified in the upper panel of Figure \ref{fig:corners}, where we show the best fitting model from the run where tier1-3 is analysed using a linear ephemeris. The best model, a sum of three Keplerian orbits, is woefully incorrect for the middle section. This shows how not only are Keplerian LTTE models not successful at predicting future eclipse times, but they are unsuccessful at interpolating. 

In the future, a better approach might be to use Gaussian Processes \citep{rasmussen_gaussian_2006} to model the shorter eclipse timing variation signals. These tools are particularly good at modelling quasi-periodic functions to stellar activity for instance \citep[in photometry and spectroscopy;][]{barros_improving_2020,faria_uncovering_2016} and would seem appropriate in the case of HW Vir.

\begin{figure*}
    \centering
    \includegraphics[width=\columnwidth]{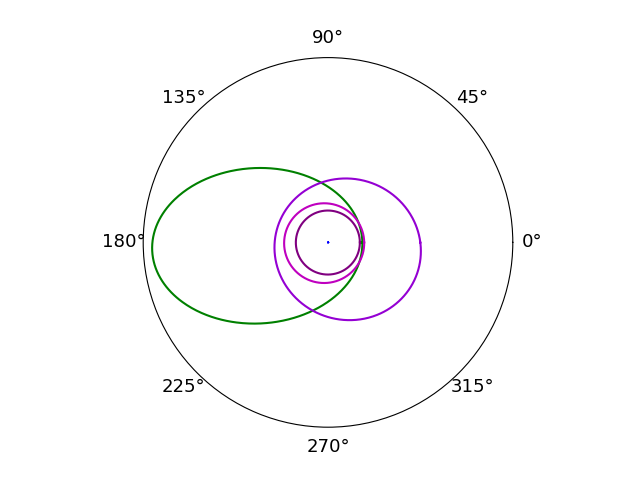}
    \includegraphics[width=\columnwidth]{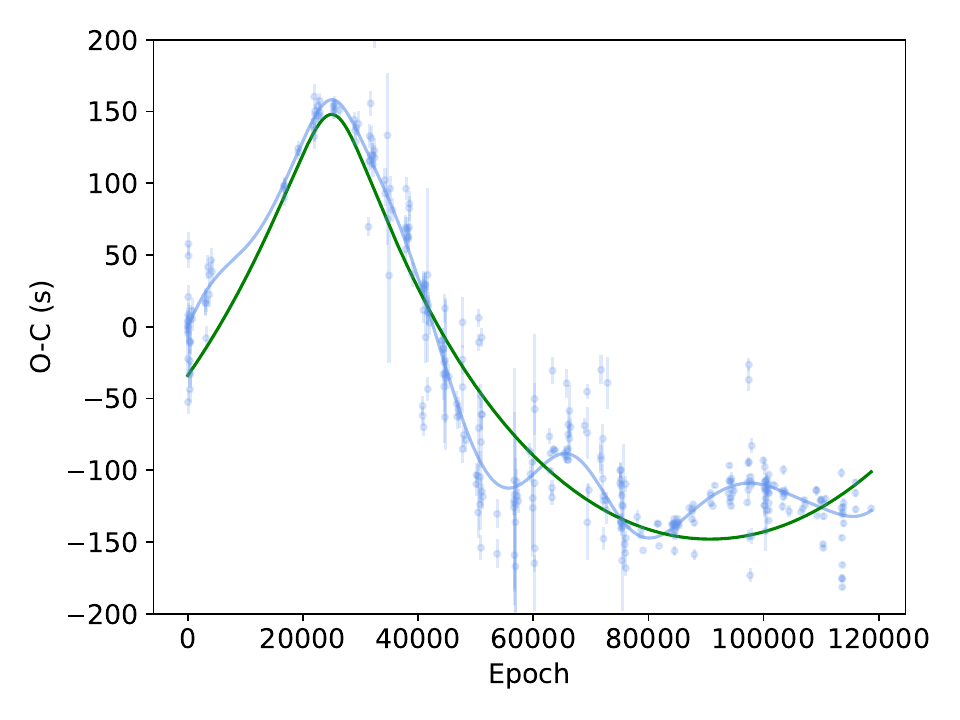}\\
    \includegraphics[width=\columnwidth]{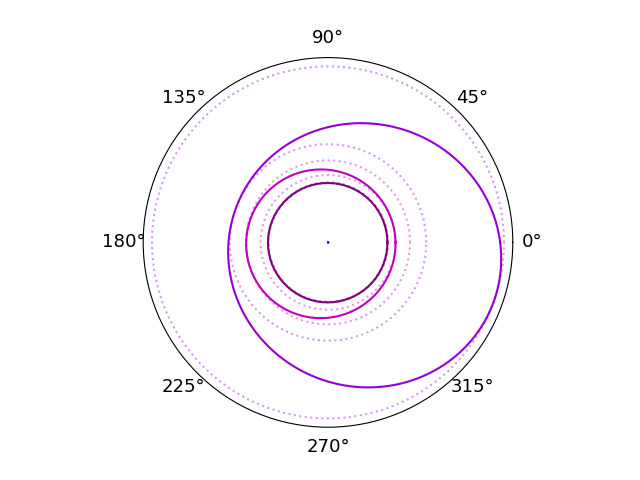}
    \includegraphics[width=\columnwidth]{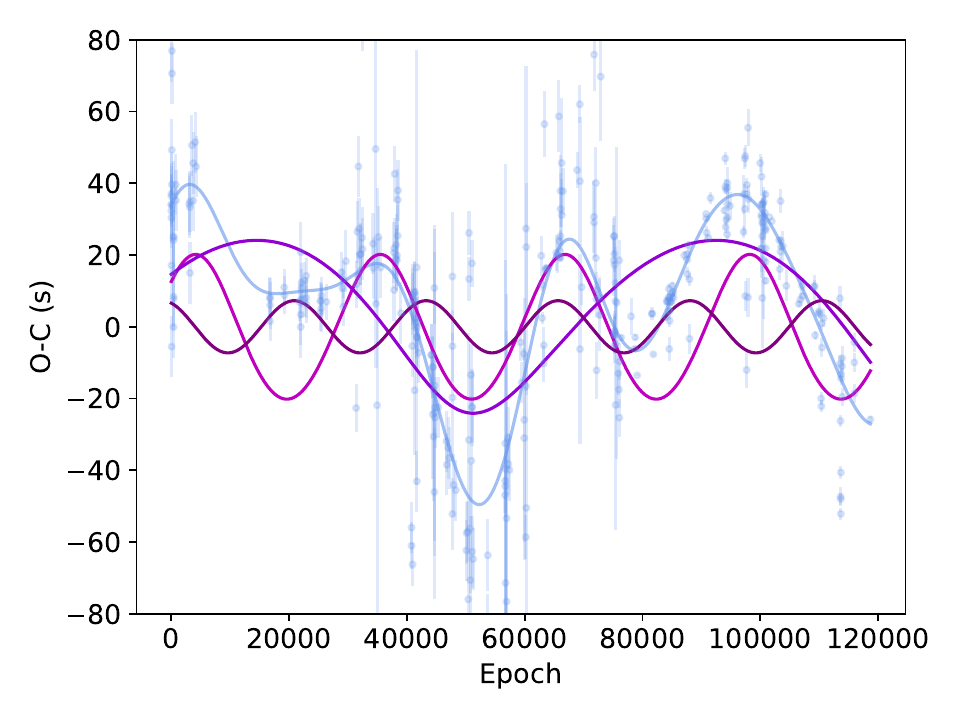}
    \caption{Orbital configurations shown on the left. Top-left: all four signals from the best-fitting model obtained from the analysis of the whole eclipse timing dataset. Bottom-left: The eccentric outer orbit is removed and the remain three signals are shown along with the dashed lines showing circular orbits at apo- and peri-centre. Best-fitting model obtained from the analysis of the whole eclipse timing dataset as well as the data itself is shown on the right (ephemeris removed). Top-right: all four signals included, the most massive is shown in green and the sum of all four in blue, this signal in green is the one we claim as the most likely candidate for being a planet. The x-axis is the Epoch \(E\) as in equation \ref{eq:etv_fit}. Bottom-right: the inner three Keplerian functions are shown with the sum of these three in blue. The data is represented with the fourth, large-amplitude, signal removed. These three signals are most likely not of planetary nature.}
    \label{fig:orbits}
\end{figure*}

\subsubsection{Results from analysing the full dataset}\label{sec:results_full}

We now show the results from an analysis of the full dataset. We allow {\tt kima} to fit freely up to $N_{\rm p} = 6$ signals along with a quadratic ephemeris. One advantage of using {\tt kima} is its ability to assess the number of signals present using Bayesian model comparison. In this case a four-signal solution is favoured as it is the highest number of planets with a significant Bayes Factor over a model with one fewer planet. The respective Bayes Factors can be seen in Table \ref{tab:bfs}. 

Four signals is more than most other analyses which only find up to two signals. The two signals already discussed (around 2500 and 4000 days) are both present in the best-fitting solution. We know this because a large fraction of the posterior sample congregate at these two orbital periods (as shown in the upper panel of Figure \ref{fig:periodogram}). The other two signals are not nearly as well constrained and do not correspond to any clear over-density in the posterior, likely because these are longer signals that have not had the chance to repeat yet, making their parameters uncertain.

Past analyses have regularly identified a signal corresponding to the one we find around 4000 days \citep[e.g.][]{beuermann_quest_2012,esmer_revisiting_2021}, however none have identified a signal near 2500 days.

\begin{table}
    \centering
    \caption{Bayes Factors produced by {\tt kima} when analysing the entire eclipsing timing dataset.}
    \begin{tabular}{c|c}
        \hline
        Number of planets compared & Bayes Factor \\
        \hline
        1 : 0 & \(>1.8\times10^{308}\) \\
        2 : 1 & \(1.8\times10^{80}\) \\
        3 : 2 & \(4.0\times10^{44}\) \\
        4 : 3 & \(1.1\times10^5\) \\
        5 : 4 & \(2.1\) \\
        6 : 5 & \(1.5\) \\
        \hline
    \end{tabular}
    \label{tab:bfs}
\end{table}

\begin{table}
    \centering
    \caption{Parameters from the analysis of the full dataset, with a quadratic ephemeris. The keplerian parameters for each signal is shown as if it was keplerian LTTE orbit. \(^{\rm a}\) for the circular orbit we combine the \(\omega\) and \(\phi\) parameters together since otherwise they are extremely correlated and no information can be gained. \(^{\rm b}\) for the two outer signals, the posterior density is not well constrained so clusters around the best-fitting solutions are used for some of the parameters. We note the uncertainties are likely too small to represent the true uncertainty in the model. \(^{\rm c}\) For these periods and amplitudes, the uncertainty is reported as the difference between the value of the parameter when a linear ephemeris is fit and when a quadratic ephemeris is fit. The median of the posterior density cluster from the quadratic ephemeris fit is retained as the quoted value. The mass distribution is the propagated in a monte-carlo way.}
    \begin{tabular}{l|l|r}
    \hline
    Parameter & Value & Units\\
    \hline
    \textit{Assumed parameters} \\
    \(M_0+M_1\)  & \(0.54\pm0.0089\) & \({\rm M_{\odot}}\)\\
    \hline
    \textit{Binary parameters} \\
    \(P_0\) & \(0.116719590^{+1.0{\rm e}-8}_{-1.8{\rm e}-8}\)  & days \\
    \(\dot{P}_0\)  & \(-1.03{\rm e}-11^{+4.4{\rm e}-12}_{-2.0{\rm e}-12}\) & days/day \\
    \hline
    \textit{Keplerian parameters} \\
    \\
    \(P_1\)  & \(2\,612^{+43}_{-41}\) & days \\
    \(K_1\)  & \(7.66^{+0.60}_{-0.81}\) & s \\
    \(e_1\) & \(<0.1\) &  \\
    \(\omega_1 + \phi_{0,1}\,^{\rm a}\)  & \(1.97\pm0.38\) & rad \\
    \(M_1(\sin i_1)\)  & \(2.88^{+0.22}_{-0.29}\) & M\(_{\rm Jup}\) \\
    \\
    \(P_2\)  & \(3\,710^{+58}_{-76}\) & days \\
    \(K_2\)  & \(21.2^{+1.5}_{-1.2}\) & s \\
    \(e_2\) & \(0.089^{+0.035}_{-0.038}\) &  \\
    \(\omega_2\)  & \(1.74^{+0.54}_{-0.44}\) & rad \\
    \(T_{{\rm per},2}\)  & \(2\,442\,500^{+380}_{-350}\) & BJD \\
    \(M_2(\sin i_2)\)  & \(6.34^{+0.45}_{-0.36}\) & M\(_{\rm Jup}\) \\
    \\
    \(P_3\,^{\rm c}\)& \(8\,400\pm1600\) & days  \\
    \(K_3\,^{\rm c}\) & \(23\pm10\) & s  \\
    \(e_3\,^{\rm b}\) & \(<0.45\) &  \\
    \(\omega_3\,^{\rm b}\) & \(4.5\pm1.3\) & rad  \\
    \(T_{{\rm per},3}\,^{\rm b}\) & \(2\,443\,600 \pm 2000\) & BJD  \\
    \(M_3(\sin i_3)\,^{\rm c}\) & \(4.0\pm1.9\) & M\(_{\rm Jup}\)  \\
    \\
    \(P_4\,^{\rm c}\) & \(15\,600\pm3500\) & days  \\
    \(K_4\,^{\rm c}\) & \(148\pm45\) & s  \\
    \(e_4\,^{\rm b}\) & \(0.6867\pm0.013\) &  \\
    \(\omega_4\,^{\rm b}\) & \(1.77^{+0.06}_{-1.4}\) & rad  \\
    \(T_{{\rm per},4}\,^{\rm b}\) & \(2\,433\,100^{+140}_{-120}\) & BJD  \\
    \(M_4(\sin i_4)\,^{\rm c}\) & \(17.4^{+6.5}_{-5.7}\) & M\(_{\rm Jup}\) \\
    \hline
    \textit{Other fit parameters} \\
    Jitter & \(0.92\pm0.23\) & s  \\
    \(T_0\) &\(2\,445\,730.556669\) & BJD  \\
    \end{tabular}
    \label{tab:parameters}
\end{table}

We do not consider any formal stability arguments. Many previous studies have found that multiple-planet solutions are unstable, and since we have a strong reason to doubt that the detected signals within the eclipsing times are produced by an orbiting planet, we feel a stability analysis is meaningless. The upper-left diagram of Figure \ref{fig:orbits} shows the orbital configuration of planets corresponding to all four signals. The inner two of them are reasonably circular, the outer two are more eccentric, and the outermost crosses the other orbits. Clearly not all four signals can be from orbiting bodies. Ignoring the outermost, eccentric orbit, the lower-left diagram in Figure \ref{fig:orbits} shows the orbital configuration of planets corresponding to the inner three signals. While these three signals do not cross into each others orbits, they present a very compact configuration, that would likely not be stable either. 

The astrometric tangential velocity implies it is more likely than not there is one orbiting companion to the HW Vir binary. Of the four signals, if one is of planetary nature, we favour the fourth and outermost signal. The analysis reported in section \ref{sec:results_chunked} casts strong doubts on the inner two signals since they appear only quasi-periodic. The third signal has too long a period to assess the consistency with the chunking method, but its 'orbital parameters' are similar to the inner two signals with a small amplitude and mild eccentricity. Compared to all others, the outermost signal lies closest to where the median value of the proper motion anomaly predicts (the dark line on Figure \ref{fig:periodogram}). 
We note that this candidate planet signal is of a similar mass and period to components of the solutions by \citet{esmer_revisiting_2021} and \citet{brown-sevilla_new_2021}. These all likely correspond to the same signal but vary in orbital period due to the data having not covered multiple cycles yet.

The plots on the right hand side of Figure \ref{fig:orbits} show the model curves for each of the signals along with the combined model and data. The upper-right panel shows the full model in the background and the outer orbit Keplerian signal in green, the lower-right panel shows the other three individual signals in shades of purple as well as their sum in the background. The three signals shown together in the lower right panel are those we claim to be most likely not produced by a planet (especially the two at shorter periods), these might be better modelled together as a Gaussian Process.

While we know that this four-component solution cannot correspond to four orbiting companions, to allow future comparison with our work, we still report the parameters of the orbits as if they were real. The parameters are detailed in Table \ref{tab:parameters}. The uncertainties associated with the parameters of the inner two orbits are well-defined as they are associated with clear clusters of posterior density (clustering using HDBSCAN is also used here). The outer two orbits do not belong to large clusters, so while they can be associated with clusters found by HDBSCAN the uncertainty on the parameters extracted from these are likely underestimated. This is because there has not been enough data for the signal to repeat. In the case of the outer signal the data has not even covered a whole phase yet. This also causes a degeneracy between the orbital parameters of the outer signals and the ephemeris terms. To partly address the underestimation, we take two analysis runs with {\tt kima}, one using a linear ephemeris and one using a quadratic. The uncertainty on the amplitude, period are then taken as the difference between the values from the two models, with the quoted value remaining the value from the analysis with a quadratic ephemeris. This is to keep the whole table representing a coherent solution. Corresponding planetary masses for the outer two signals are then produced in a monte-carlo way. It should be noted this is therefore not a statistically derived uncertainty, but is a rough representation of the uncertainty from the fitting procedure. 

\section{Discussion and Conclusion}\label{sec:disconc}

Our analysis of the eclipse times of HW Vir does not find a single conclusive solution. This is in agreement with past work  since every published solution has subsequently diverged from new data acquired afterwards \citep[e.g.][]{pulley_eclipse_2022}. We have shown that there are two strong periodicities in the full dataset which are also seen independently in some of the smaller \textit{chunks}. However, though signals can be found near these periods in most of the \textit{chunks}, their posterior distributions in \(P\) and \(K\) are not completely coherent in time, nor statistically consistent with one another. While past analyses have identified the periodicity around 4000 days, none have identified the signal around 2500 days.

We have presented our solution from the analysis of the whole eclipse timing dataset performed in a fully Bayesian way using nested sampling within {\tt kima}. This 4-component model includes signals at both of the strong periodicities. It is abundantly clear that not all four of the signals in this model are due to orbiting companions, in fact it is possible that none of them are. There likely must be some other mechanism involved for the variation in eclipse times, one possibility being a magnetic effect which is not yet fully understood \citep[e.g.][]{lanza_internal_2020}. We suggest that using a Keplerian prescription for non-planetary, quasi-periodic signals like what these appear to be is insufficient and that using a Gaussian Process method may work better \citep[as in][]{faria_uncovering_2016}. We propose that of the four signals, if one is produced by a planet, it is most likely to be the outermost one. This signal also best fits the astrometric evidence and has a signature that looks most different to the other three.

We have applied the proper motion anomaly method to V471 Tau and confirmed the non-detection of a previously proposed orbiting brown dwarf. We have also applied it to HW Vir and shown that there is a tentative 2-\(\sigma\) signal of an acceleration due to an orbiting body. From the upper limit this poses, we can discount some previously proposed companions which are too massive to be consistent with the proper motion anomaly. Comparing the astrometric signal with the four signals we extract from the eclipse timings in Fig.~\ref{fig:periodogram}, we find that the outermost signal is the most consistent. If correct this corresponds to a $17~\,\rm M_{jup}$, $16\,000~\rm day$, highly eccentric companion. Thanks to additional data, a longer baseline, and an improved astrometric solution, the full epoch astrometry from \textit{Gaia} will (circa 2025) likely be able to help resolve whether the HW Vir binary is indeed host to an orbiting circumbinary companion.The \textit{Gaia} baseline will still be much shorter than the most likely planet's orbital period, so while the whole period would not be covered, astrometry may still tell us whether or not such a planet exists, independently of the ETVs. This will help identify which (if any) of the varying eclipse timing signals is actually caused by that orbiting body. In turn, this will help isolate the functional form of the potential new physics causing the other signals (for instance the 2500 and 4000 day signals). As described in \citet{sahlmann_gaias_2015}, thanks to {\it Gaia}'s final solution, other post-common envelope circumbinary systems will be solved astrometrically with our paper being the first attempt at doing so.

\section*{Acknowledgements}

This paper is dedicated to the memory of Tom Marsh whose  kindness, curiosity and many discussions inspired much  interest in circumbinary planets, and more specifically to those in post-common envelope systems.

We thank Annelies Mortier and Lalitha Sairam for productive discussions and insightful comments on this work.

This research leading to these results is supported by two grants from the European Research Council (ERC) under the European Union's Horizon 2020 research and innovation programme (grant agreements n$^\circ$ 803193/BEBOP and 951549/UniverScale). 

The computations described in this paper were performed using the University of Birmingham's BlueBEAR HPC service.

\section*{Data Availability}
 
 Proper motion anomaly and eclipse timing data are available from the original publications cited in this work and described in Sections \ref{sec:ast} and \ref{sec:data}.
The version of {\tt kima} used to analyse the eclipse timing data is available at https://github.com/TomAB99/kima/tree/etvs



\bibliographystyle{mnras}
\bibliography{HWVir} 








\bsp	
\label{lastpage}
\end{document}